\documentclass[a4paper,11pt]{article}

\pdfoutput=1 
\usepackage{jinstpub} 

\title{\boldmath Study of neutron irradiation effects in Depleted CMOS detector structures  $^{*}$ \note[*]{The work was partly done in the framework of the RD50 collaboration.}}

 \author[a,1]{I.~Mandi\' c \note{Corresponding author.}}

 \author[a]{V. ~Cindro}
 \author[a]{J. ~Debevc}
 \author[a]{A. ~Gori\v sek}
 \author[a]{B. ~Hiti}
 \author[a]{G. ~Kramberger}
 \author[a]{P. ~Skomina}
\author[a]{M. ~Zavrtanik}
\author[a,b]{M. ~Miku\v z}
\author[c]{E. ~Vilella}
\author[c]{C. ~Zhang}
\author[c]{S. ~Powell}
\author[c*]{M. ~Franks}
\author[d]{R. Marco-Hernandez}
\author[e]{H. Steininger}

\affiliation[q]{Jo\v zef Stefan Institute, Jamova 39,Ljubljana, Slovenia}
\affiliation[b]{University of Ljubljana,Faculty of Mathematics and Physics, Jadranska 19, Ljubljana, Slovenia}
\affiliation[c]{University of Liverpool, Department of Physics, Liverpool, UK}
\affiliation[*]{now FBK, Trento, Italy}
\affiliation[d]{IFIC (CSIC-UV), Valencia, Spain}
\affiliation[e]{HEPHY, Vienna, Austria}

\emailAdd{igor.mandic@ijs.si}

 \abstract{In this paper the results of Edge-TCT and I-V measurements with passive test structures made in LFoundry 150 nm HV-CMOS process on p-type substrates with different initial resistivities ranging from 0.5 to 3 k$\Omega$cm are presented. Samples were irradiated with reactor neutrons up to a fluence of 2$\cdot$10$^{15}$ n$_{\mathrm{eq}}$/cm$^2$. The depletion depth was measured with Edge-TCT. The effective space charge concentration $N_{\mathrm{eff}}$ was estimated from the dependence of the depletion depth on bias voltage and studied as a function of neutron fluence. The dependence of $N_{\mathrm{eff}}$ on fluence changes with initial acceptor concentration in agreement with other measurements with p-type silicon. A long term accelerated annealing study of $N_{\mathrm{eff}}$ and detector current up to 1280 minutes at 60$^\circ$C was made. It was found that $N_{\mathrm{eff}}$ and current in reverse biased detector behave as expected for irradiated silicon.}

\keywords{
  Particle tracking detectors (Solid-state detectors), Radiation-hard detectors, DMAPS}

\begin{document}
\maketitle
\flushbottom


\section{Introduction}
\label{intro}

Monolithic charged particle detectors in depleted CMOS technology offer several advantages over hybrid detectors, such as higher granularity, lower detector mass, simplified assembly, production in industrial process on large wafers in high volume foundries, etc. Usage of high voltage CMOS technology enables charge collection from the depletion volume and therefore higher speed and radiation hardness necessary for operation in a hadron collider environment. Development of depleted CMOS detectors for hadron colliders has been very intense in recent years and is still keeping its momentum  \cite{peric, Wermes, Pernegger,Dyindal}. RD50 collaboration \cite{RD50} has joined the efforts and produced two detector prototypes in 150 nm LFoundry HV-CMOS process \cite{LFoundry} containing several test structures for various radiation hardness studies. In this work, measurements with passive pixel detector structures on RD50-MPW1 and RD50-MPW2 chips are described.

\section{Samples}

\begin{figure}[!hbt]
\centering
\includegraphics[width=1.1\textwidth]{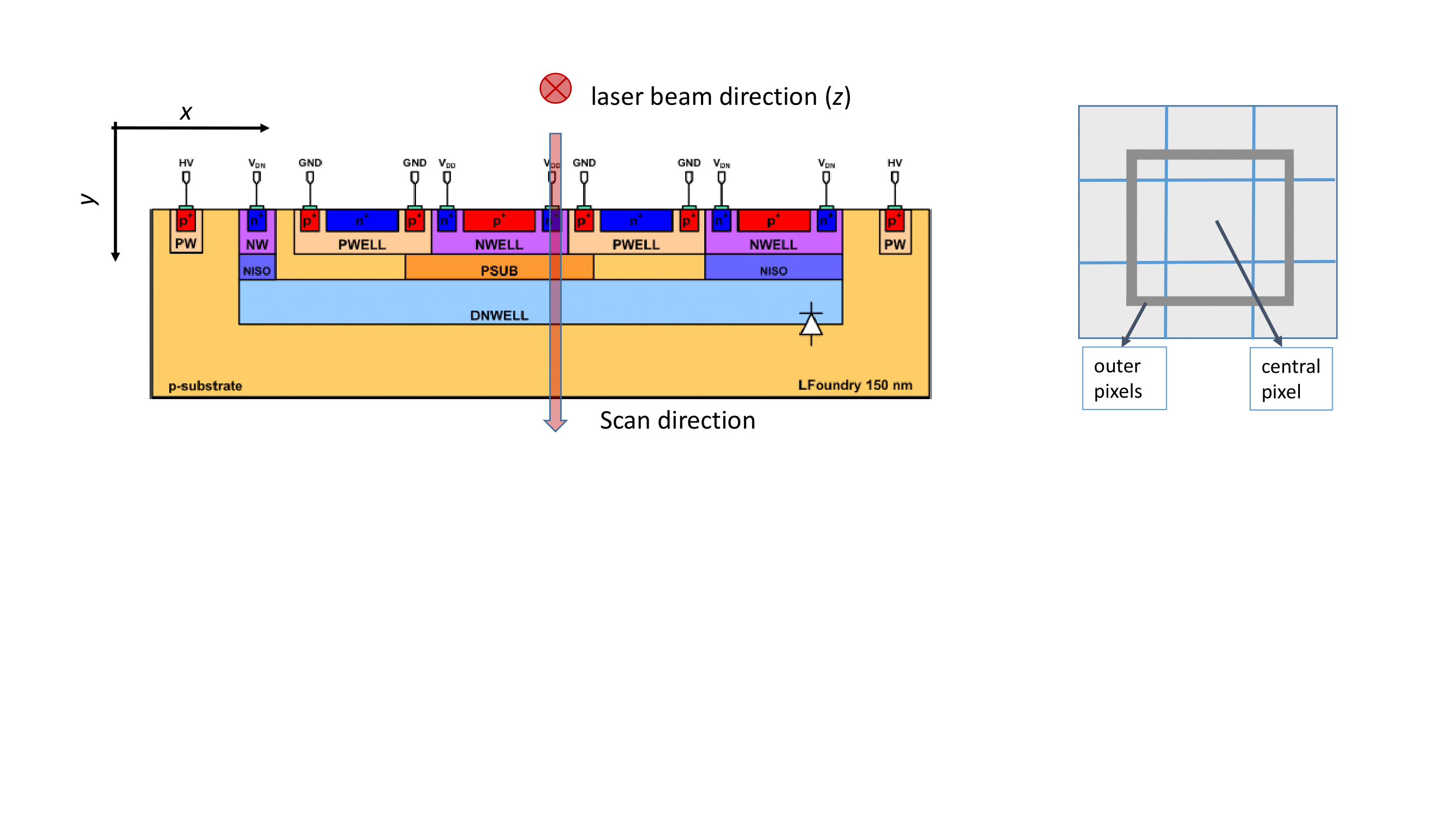} \\
  \vspace{-4 cm} 
   \hspace{4.5cm} a)  \hspace{6cm} b) \\ 
   \vspace{5mm}

   \caption{Figure a) shows a simplified scheme of a HV-CMOS pixel. The definition of coordinate system can be seen and the direction of the laser beam scan in E-TCT  is shown. Figure b) represents the 3x3 pixel array in which DNWELLs of the outer 8 pixels are connected to one bond-pad and DNWELL of the central pixel is contacted to a separate bond pad. The substrate is contacted via the PW implants, all connected to the same bond pad.  } 
\label{schemes1}
\end{figure}

\begin{figure}[!hbt]
\
\centering
\includegraphics[width=1.1\textwidth]{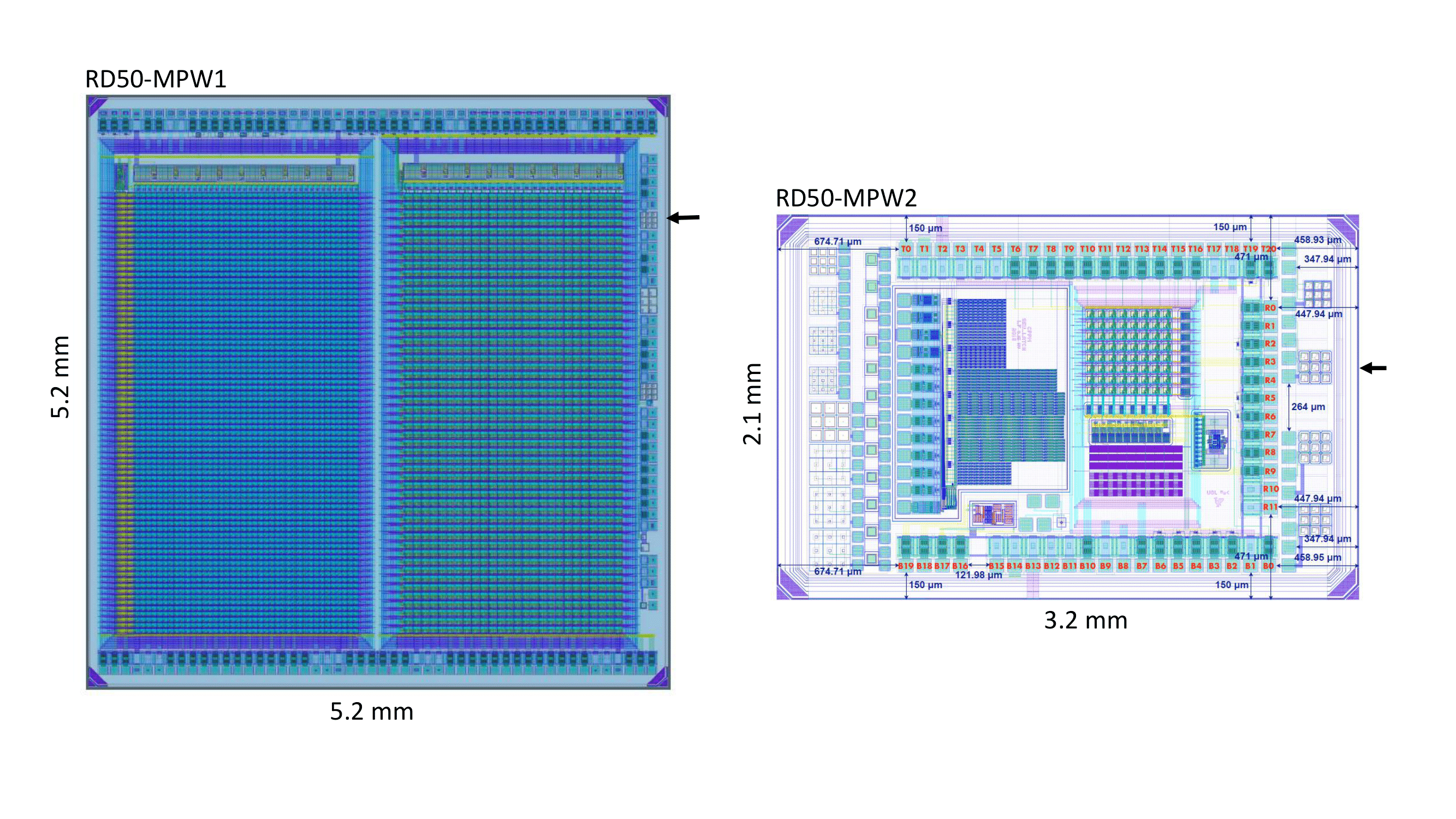} \\

   \vspace{-9 mm} 
      \hspace{2cm} a)  \hspace{7cm} b) \\ 
      \vspace{5mm}
\caption{a) Layout of the RD50-MPW1 chip and b) RD50-MPW2 chip. In figure b) 3x3 pixel structures can be seen near the right edges of the chips. Arrows point to the arrays used in this work.}
\label{MPW-schemes}
\end{figure}

As the name suggests, RD50-MPW1 was the first chip designed by the RD50 collaboration and details about its functionalities and performance can be found in \cite{EvaVertex_2019}. Due to imperfections in the design of the RD50-MPW1 chip, pixel current was higher than expected and already started to rise rapidly above a reverse bias voltage of $\sim $ 30 V. This initiated the submission of RD50-MPW2 \cite{Vertex_2020Ricardo} with the shortcomings corrected so taht the pixels could be biased with over 110 V.

The designs of the pixels and pixel arrays are very similar in RD50-MPW1 and RD50-MPW2. In RD50-MPW1 the pixels are 50 $\mu$m $\times$ 50 $\mu$m and the spacing between PW and NW implants (see Figure  \ref{schemes1}a) is 3 $\mu$m while in RD50-MPW2 the dimensions are 60 $\mu$m $\times$ 60 $\mu$m with an inter-electrode spacing of 8 $\mu$m. Larger spacing is one of the main reasons for better breakdown performance of structures on MPW2.

Chips were produced on p-type substrates with different initial resistivities. In this work measurements with chips from RD50-MPW1 and RD50-MPW2 are described and the samples are listed in table \ref{table1}.
Figure \ref{schemes1}a) shows a simplified cross section of a HV-CMOS pixel. DNWELL is the charge collecting electrode covering a large part of the pixel cross section. Shallow wells above DNWELL, named PWELLs and NWELLs, contain analogue readout electronics in active pixels.  These include a high
impedance circuit to bias the collecting electrode, a charge sensitive amplifier and a CMOS comparator with a 4-bit DAC to
locally tune small threshold voltage variations due to offset (for more detail please see \cite{EvaVertex_2019}). In a passive test structure, shallow wells are left without any elements. In this work only passive pixels were used with an external amplifier connected to DNWELL. Measurements were made with 3x3 passive pixel arrays. 
In these arrays, DNWELLs of the outer 8 pixels are connected to one bond-pad and DNWELL of the central pixel is contacted to a separate bond pad as sketched in Figure \ref{schemes1}b). The substrate is contacted via the PW implants.

Figure \ref{MPW-schemes}a) shows the layout and dimensions of the RD50-MPW1 and Figure \ref{MPW-schemes}b) of RD50-MPW2 chips containing various test structures.  Arrows in Fig. \ref{MPW-schemes} point to the 3x3 arrays used in this work. The structures are placed near the edge of the chip to enable Edge-TCT measurements.

\begin{table}
\centering
\begin{tabular}{c c c} 
Wafer & Chip version & Initial resistivity \\
 & & [k$\Omega$cm] \\
 \hline 
W8 &  RD50-MPW2 & 0.5 \\
W9 & RD50-MPW1 & 0.7  \\
W10 & RD50-MPW1 & 1.3 \\
W14 & RD50-MPW2 & 2.2 \\
\hline 
\end{tabular}
\caption{List of samples and initial resistivities of the substrates}
\label{table1}

\end{table}
 
\section{Irradiation}

Samples were irradiated with neutrons in the TRIGA reactor in Ljubljana \cite{Reactor1,Reactor2} to 1 MeV neutron equivalent
fluences ranging from 1$\cdot$10$^{13}$ n$_{\mathrm{eq}}$/cm$^2$ to  2$\cdot$10$^{15}$ n$_{\mathrm{eq}}$/cm$^2$. 
The maximum power of the TRIGA reactor in Ljubljana is 250 kW and at this power the 1 MeV neutron equivalent flux in the irradiation tube used for this work is 1.5$\cdot$10$^{12}$ n$_{\mathrm{eq}}$/cm$^2$/s. Irradiation to 1$\cdot$10$^{14}$ n$_{\mathrm{eq}}$/cm$^2$ therefore takes slightly over one minute. Reactor power, which is proportional to neutron flux, is reduced for irradiation to lower fluences because it is easier to handle irradiation times longer than one minute. The neutron flux in the irradiation channel was determined with a precision of about 10\% by measurements of leakage-current increase in dedicated silicon diodes \cite{dosimetry}. 

\section{Measurements}


The depletion depth was estimated with Edge TCT (E-TCT) \cite{Edge-TCT} using Particulars measurement system \cite{Particulars}. In E-TCT a pulsed narrow laser beam is directed to the side of the chip. The chip is moved in a direction perpendicular to the laser beam in 5 $\mu$m steps as sketched in Figure \ref{schemes1}a). Laser light with a wavelength of $\lambda$ = 1064 nm and absorption length of about 1 mm in silicon is used to release electron-hole pairs along the laser beam line inside the investigated test structure. If the light crosses the depleted region of a pixel, a current pulse is induced on the pixel electrodes which is observed by the E-TCT measurement system. E-TCT measurements were made with 3 x 3 pixel arrays shown in Fig. \ref{schemes1}. The central pixel was connected to the wide-bandwidth amplifier of the E-TCT system and to the high voltage using a bias-T circuit. This circuit enables DC connection of the high voltage source to the pixel electrode and AC/RF connection to the amplifier via a capacitor. To come closer to the conditions of a larger pixel array, the surrounding 8 electrodes were connected to the same high voltage potential but not to the E-TCT system amplifier. Induced currents on the central pixel were recorded by the oscilloscope and saved to the computer. The time integral of the induced current in the 10 ns time window was a measure of collected charge. 
In this work the focus is on the charge collected from the depletion region which is necessary for application in hadron collider environments. The chosen integration time is sufficient for charge collection from the depletion region and longer integration time would increase the noise on measured charge values because of oscillations, reflections and baseline fluctuations in the tail of the recorded pulses.
 With the E-TCT method, collected charge can be measured as a function of the detector depth (distance from the chip surface) providing charge collection profiles. The depleted depth is estimated from the width of the charge collection profiles measured in the scan along the edge of the chip as indicated in Fig. \ref{schemes1}a). Similar types of measurements are described in several publications \cite{JinstChess, FirstLF, hitixfab, SecondLF, cavallaro, franks} and more details can be found therein.

\begin{figure}
\centering
\begin{tabular}{c c} 
 \includegraphics[width=0.5\textwidth]{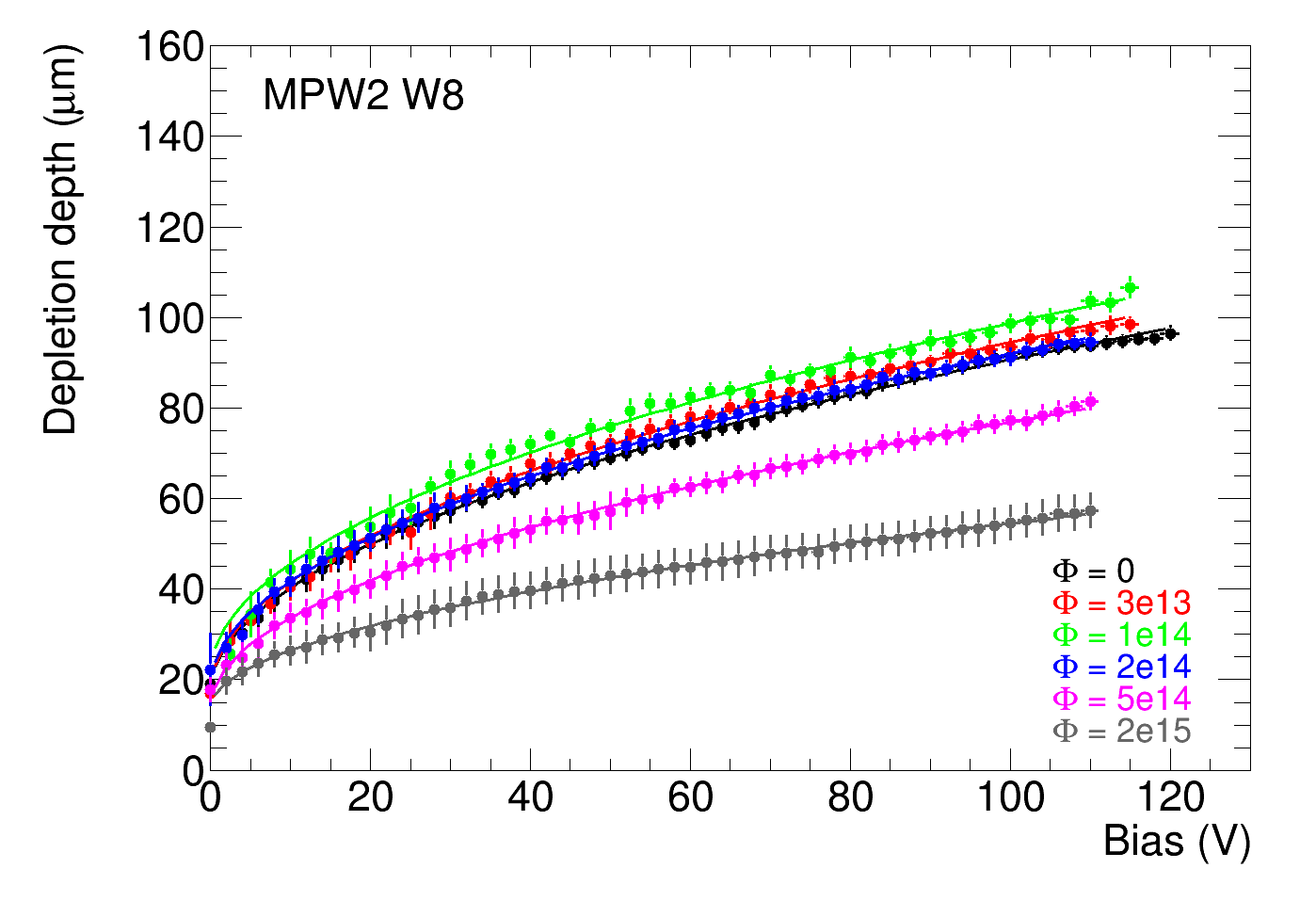} & 
 \includegraphics[width=0.5\textwidth]{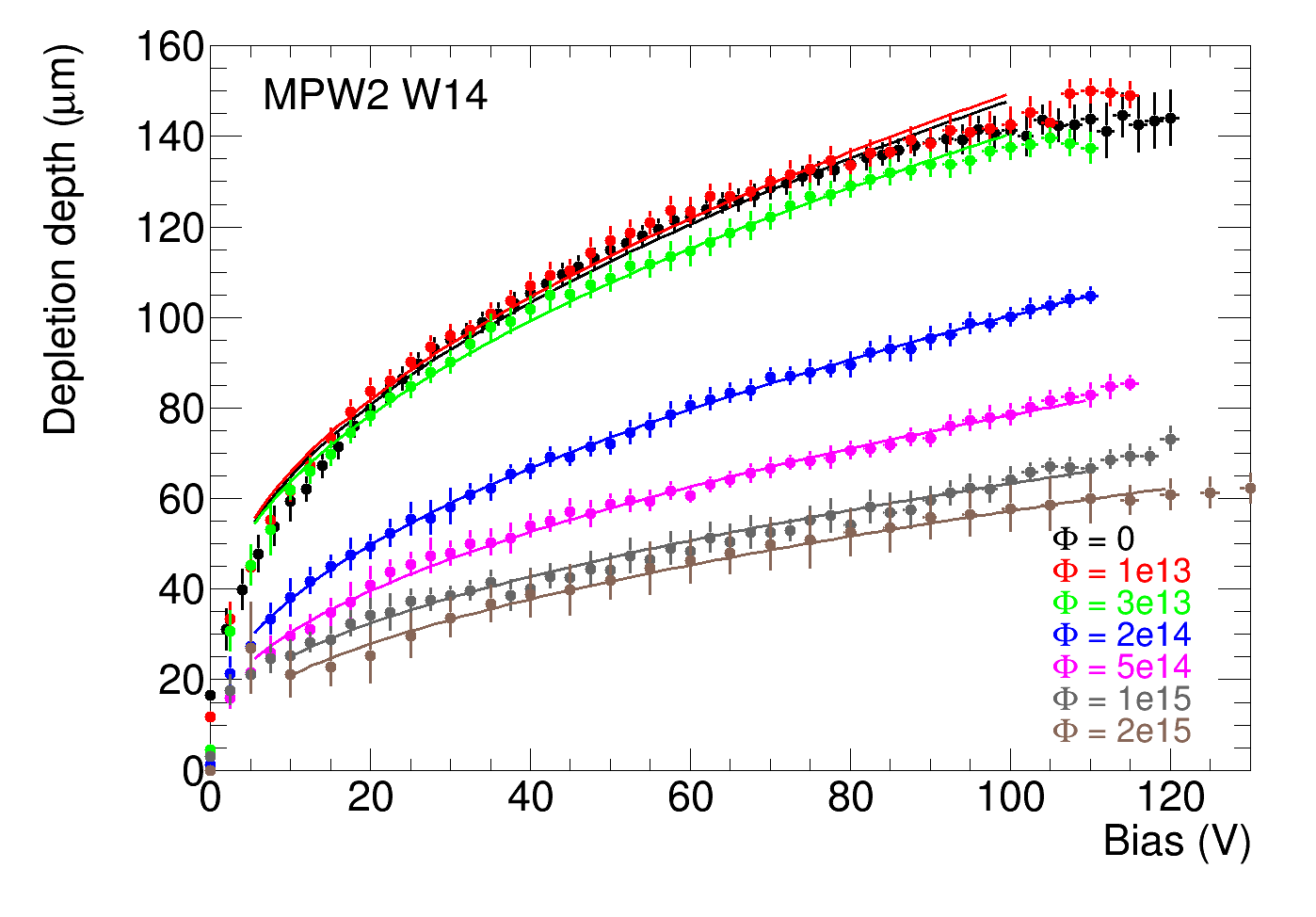} \\ 
    a) & b) \\
 \includegraphics[width=0.5\textwidth]{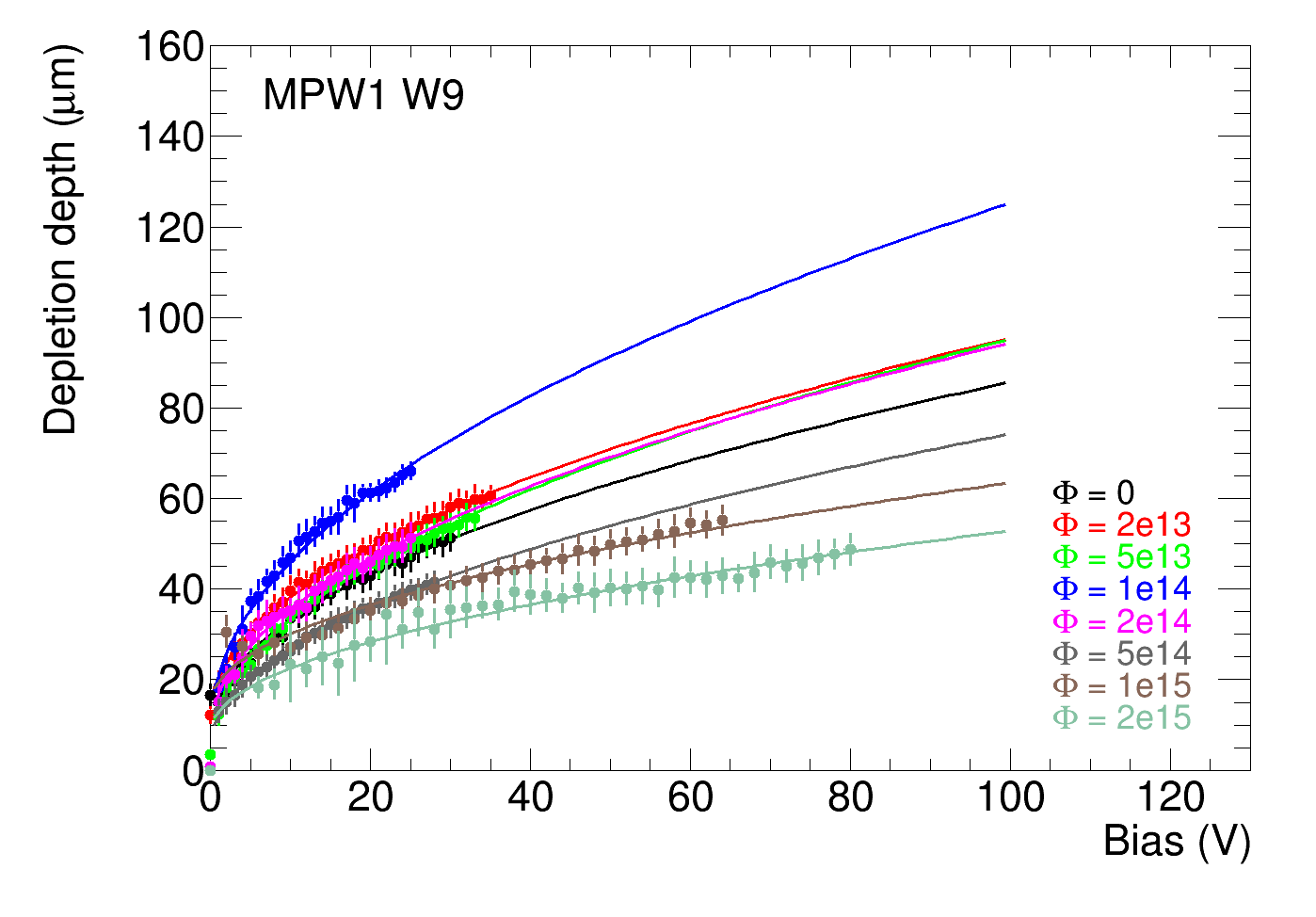} & 
 \includegraphics[width=0.5\textwidth]{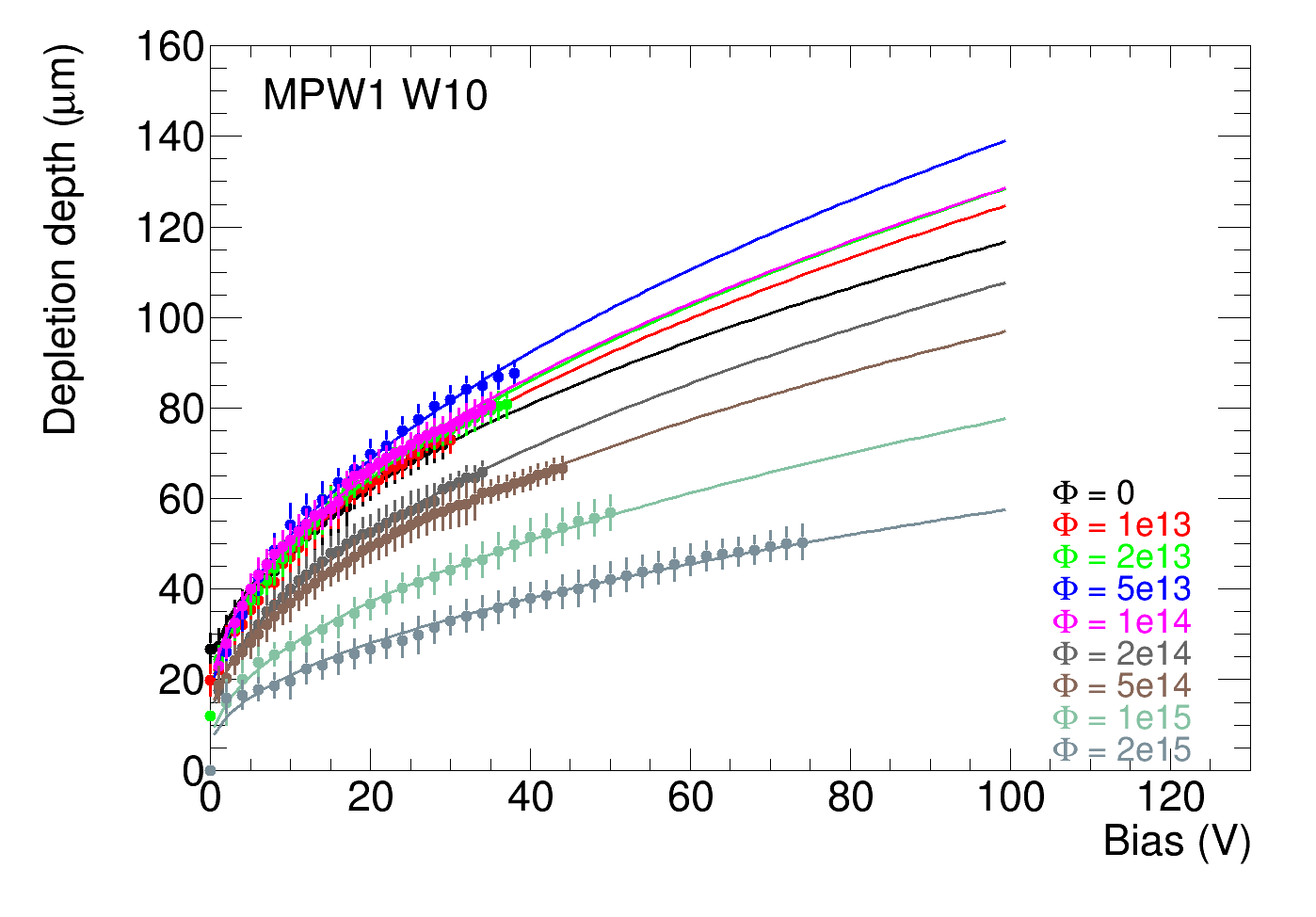} \\ 
 c) & d) \\
  \end{tabular}
\caption{Depletion depth measured as a function of bias voltage for samples from RD50-MPW1 and RD50-MPW2 chips with different initial resistivities - see table \ref{table1}: a) W8 (RD50-MPW2, b) W14 (RD50-MPW2), c) W9 (RD50-MPW1) and  d) W10 (RD50-MPW1). Measurements were made before irradiation and after irradiation to 1 MeV neutron equivalent fluences written in the graphs}
\label{deplvsbias}
\end{figure}

Figure \ref{deplvsbias} shows measurements of depletion depth vs. bias voltages for chips with different initial resistivities listed in table \ref{table1}. The graphs show measurements before irradiation and after irradiation with neutrons to equivalent fluences indicated in the figures. Before measurements samples were annealed for 80 minutes at 60 $^\circ$C. In all samples except for the highest initial resistivity (W14 in Fig \ref{deplvsbias}b), an increase of the depletion depth compared to that before irradiation can be observed. For the lower initial resistivity samples (W8 and W9) the depletion depth is larger than before irradiation up to the fluence of about $2\cdot 10^{14}$ n$_{\mathrm{eq}}$/cm$^2$, while for W10 increased depletion is measured up to $1\cdot 10^{14}$ n$_{\mathrm{eq}}$/cm$^2$. This increase is the consequence of the initial acceptor removal \cite{dosimetry, JinstChess, accrem} resulting in a lower effective space charge concentration than before irradiation. 

The lines fitted to the measured depletion depth in Fig. \ref{deplvsbias} are a result of a fit of the function \ref{fitfunc}:

\begin{equation}
  w(V_{\mathrm{bias}}) = w_0 + \sqrt{\frac{2\epsilon_r\epsilon_0}{e_0N_{\mathrm{eff}}}V_{\mathrm{bias}}}
  \label{fitfunc}
\end{equation}

where $w$ is the depletion depth, $N_{\mathrm{eff}}$ is the effective space charge concentration, $V_{\mathrm{bias}}$ the bias voltage, $e_0$ the elementary charge, $\epsilon_0$ the dielectric constant and $\epsilon_r$ the relative permittivity of silicon.
Parameter $w_0$ is introduced to account for effects of finite laser beam diameter, intrinsic depletion etc. contributing to depletion measured at 0 V bias voltage. Parameters $w_0$ and $N_{\mathrm{eff}}$ are free and their values are extracted from the fit. 
Values of parameter $w_0$ vary between 5 and 25 $\mu$m. Every fluence point was measured with different chip and the value of $w_0$ depends on the details of sample mounting and laser light focus etc.
It can be seen in Fig \ref{deplvsbias} that the function \ref{fitfunc} can be fitted to measurements relatively well with some 
discrepancies seen at low voltages for W14 (\ref{deplvsbias}b)) for lowest fluences. This effect is not fully understood but it indicates that describing all of the effects contributing to the measured depth at low voltages by a single constant $w_0$ is not sufficient. But function \ref{fitfunc} fits the data over a wide interval of bias voltages so it can be used to estimate $N_{\mathrm{eff}}$.   
 
Dependence of $N_{\mathrm{eff}}$ on neutron fluence can be described with \cite{JinstChess}: 
 
\begin{equation}
N_{\mathrm{eff}} = N_{\mathrm{eff,0}} - N_{\mathrm{c}} \cdot (1 - e^{-c\Phi_{\mathrm{eq}}}) + g_c \cdot \Phi_{\mathrm{eq}}
\label{neffvsphi}
\end{equation}
where $N_{\mathrm{eff,0}}$ is the initial acceptor concentration of the substrate, $N_{\mathrm{c}}$ the concentration of the removed
acceptors, $c$ the removal constant and $g_c$ the introduction rate of stable deep acceptors for irradiation with neutrons.

\begin{figure}
\centering
\begin{tabular}{c c} 
 \includegraphics[width=0.5\textwidth]{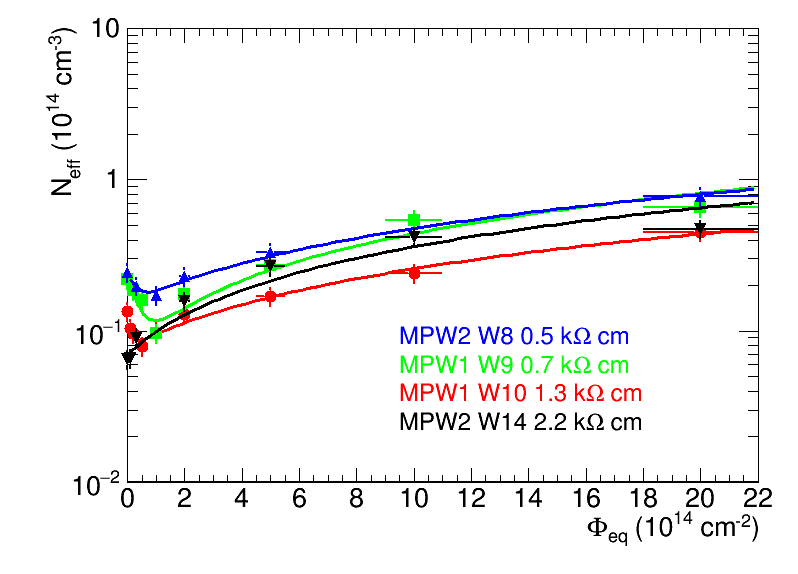} & 
  \includegraphics[width=0.5\textwidth]{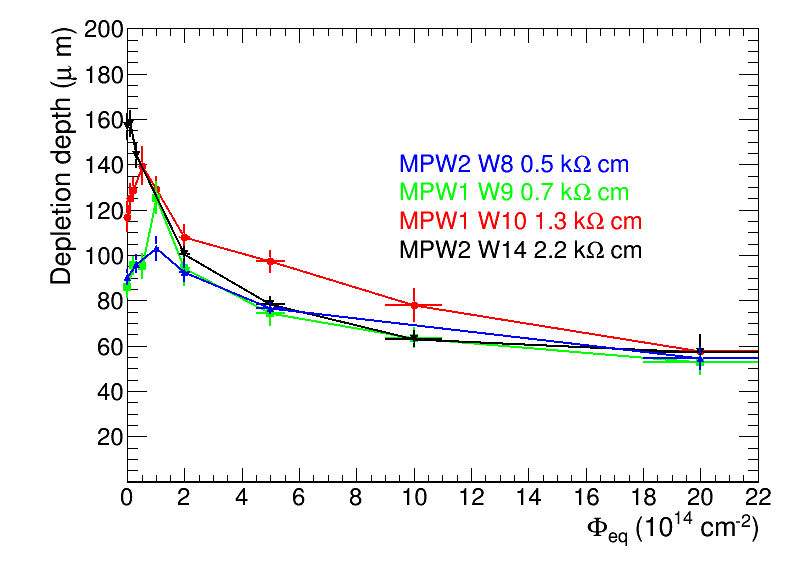} \\ 
    a) & b) \\
  \end{tabular}
\caption{Figure a) measured $N_{\mathrm{eff}}$ (points) and result of the fit (lines) of equation \ref{neffvsphi} as a function of fluence. Figure b) shows depletion depth measured at 100 V as a function of fluence for the four samples.}
\label{Neff vs fluence}
\end{figure}

Figure \ref{Neff vs fluence}a) shows the evolution of effective space charge concentration on fluence with the theoretical curve described in equation \ref{neffvsphi} fitted to measured points. A decrease of  $N_{\mathrm{eff}}$ after the first fluence steps can clearly be seen for the three lower initial resistivities followed by the increase at the same rate for all samples after higher fluences. Similar behaviour was measured in several other experiments \cite{FirstLF, SecondLF, franks}. Table \ref{table2} summarizes the parameters obtained from the fit of function \ref{neffvsphi}. 
There are large uncertainties but the values are not in contradiction with the literature \cite{FirstLF, franks, removal,KrambiHiroshima} stating that removal constant $c$ is higher for higher initial resistivities. Stable damage introduction rate $g_c$ is somewhat larger than typical for neutron irradiated silicon but this is consistent with similar types of measurements \cite{FirstLF, SecondLF, franks}.  In table \ref{table2} only statistical errors calculated in the fitting procedure are listed. Systematic uncertainties of the parameters were not evaluated although they might be significant. The aim of this work was to check if the behaviour of the substrate after irradiation is roughly within expectations. For more accurate measurements of parameters in equation \ref{neffvsphi} a dedicated experiment would be needed. 

For the sample with highest initial resistivity, W14, the parameters $N_{\mathrm{c}}$ and $c$ could not be estimated in the range of fluences investigated in this work. As mentioned above, higher initial resistivity is associated with higher value of $c$ which means that acceptor removal process is finished at lower fluences. To observe acceptor removal and estimate the parameters $N_{\mathrm{c}}$ and $c$ measurements would have to be made at fluences below $1\cdot 10^{13}$ n$_{\mathrm{eq}}$/cm$^2$ with W14. Therefore, only the values of parameters $N_{\mathrm{eff,0}}$ and  $g_c$ could be estimated from the fit of equation \ref{neffvsphi}. 

The number of free carriers released by the passage of a charged particle is proportional to the depletion depth so it is one of the crucial parameters determining the detector performance. Fig. \ref{Neff vs fluence}b) shows the depletion depth measured at $V_{\mathrm{bias}} = 100$ V. RD50-MPW1 chips (W9 and W10) could not be biased with 100 V so the depletion depths were estimated from the extrapolation of the fitted function at 100 V (see Fig. \ref{deplvsbias}). A bias of 100 V was chosen for this figure because it is a realistic value at which detectors made in this technology might be operated in an application. 
One can see that there are significant differences before irradiation and at low fluences, while above $\approx 2\cdot 10^{14}$ n$_{\mathrm{eq}}$/cm$^2$ a similar depletion depth is measured in all samples.  The sample W10 exhibits a somewhat higher depletion depth at fluences between $2\cdot 10^{14}$ n$_{\mathrm{eq}}$/cm$^2$ and $1\cdot 10^{15}$ n$_{\mathrm{eq}}$/cm$^2$, however these values were obtained from extrapolation to 100 V and have therefore significant uncertainties.

\begin{table}
\centering
\begin{tabular}{c c c c c c} 
Wafer & Chip version & $N_{\mathrm{eff},0}[10^{14} cm^{-3}]$ & $N_c/N_{\mathrm{eff},0}$ & c[10$^{-14} cm^{2}$] & $g_c[10^{-2} cm^{-1}]$\\
 \hline 
W8 &  RD50-MPW2 & 0.25 $\pm$ 0.04 & 0.39 $\pm$ 0.14 & 3.5 $\pm$ 3.2 & 3.3 $\pm$ 0.6\\
W9 & RD50-MPW1 & 0.23 $\pm$ 0.03 & 0.72 $\pm$ 0.10 & 2.5 $\pm$ 0.9  & 3.9 $\pm$ 0.6\\
W10 & RD50-MPW1 & 0.13 $\pm$ 0.02 & 0.43 $\pm$ 0.10 & 8.1 $\pm$ 5.4 & 1.8 $\pm$ 0.3 \\
W14 & RD50-MPW2 & 0.07 $\pm$ 0.01 & $/$ & $/$ & 2.9 $\pm$ 0.4 \\ 
\hline 
\end{tabular}
\caption{Parameters of fit of fluence dependence of effective space charge concentration $N_{\mathrm{eff}}$ described in equation \ref{neffvsphi}. }
\label{table2}
\end{table}

\section{Annealing effects}

The measurements shown above were made with samples annealed for 80 minutes at 60$^\circ$C. The measurements were made before and after annealing and it was observed that the depletion depth increased and $N_\mathrm{eff}$ decreased confirming the beneficial effect of short term annealing. With samples from W8 and W14 irradiated to $\Phi_{\mathrm{eq}} = 2\cdot 10^{15}$ n$_{\mathrm{eq}}$/cm$^2$ a long term annealing study was made in which measurements were taken after annealing for  80, 160, 320, 640 and 1280 minutes at 60$^\circ$C.

\begin{figure}
\centering
\includegraphics[width=0.8\textwidth]{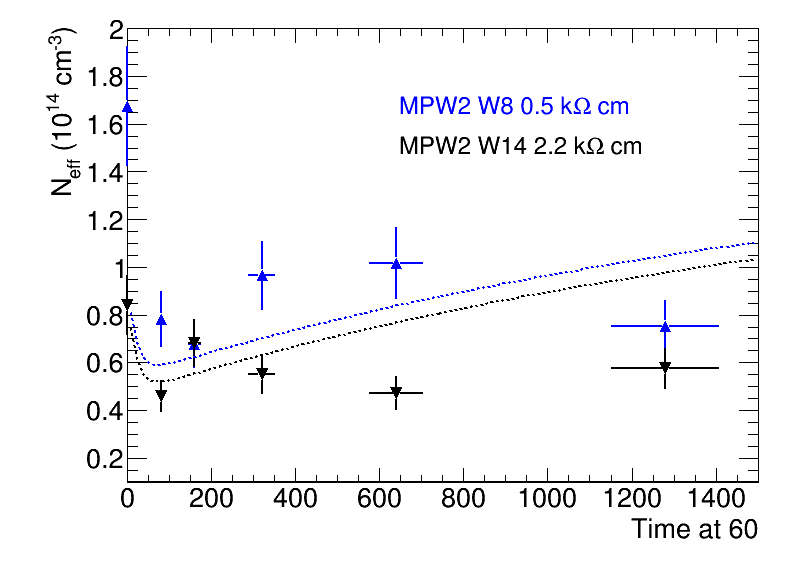} 
\caption{Effective space charge concentration $N_{\mathrm{eff}}$ measured with W8 and W14 irradiated to $\Phi_\mathrm{eq} = 2\cdot 10^{15}$ n$_\mathrm{eq}$/cm$^2$ as a function of annealing time at  60$^\circ$C. Dotted lines show annealing function with parameters measured by RD48 collaboration. }
\label{NeffAnnealing}
\end{figure}

The dependence of $N_{\mathrm{eff}}$ on annealing time is shown in Figure \ref{NeffAnnealing}. A significant effect of the first (80 minutes) annealing step can be seen, while the measured points at longer annealing times are scattered and do not exhibit a significant annealing effect. For comparison, curves of RD48 annealing model \cite{dosimetry} for neutron irradiated silicon are added in the plot.  It may be noted that the increase of effective space charge concentration at long annealing times (reverse annealing) is somewhat smaller than expected from the model. For the curves drawn in Figure \ref{NeffAnnealing}, parameters describing float zone silicon are used. It was measured in \cite{reverseanneal} that neutron irradiated diodes fabricated on magnetic Czocharlski substrate show less reverse annealing compared to diodes from float zone silicon. RD50-MPW1(and 2) chips are made on Czochralski substrate and this may be the reason for the discrepancy seen in Figure \ref{NeffAnnealing}.

\section{Leakage current}

\begin{figure}
\centering
\begin{tabular}{c c} 
 \includegraphics[width=0.5\textwidth]{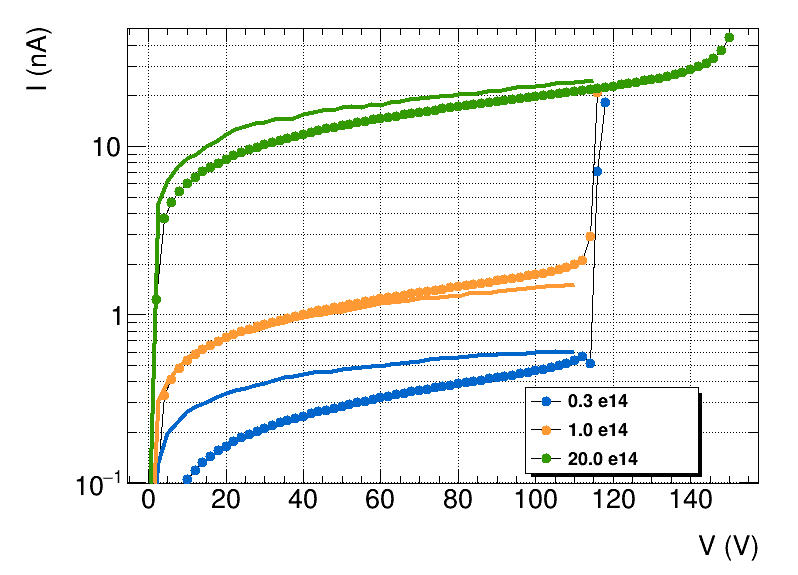} & 
  \includegraphics[width=0.5\textwidth]{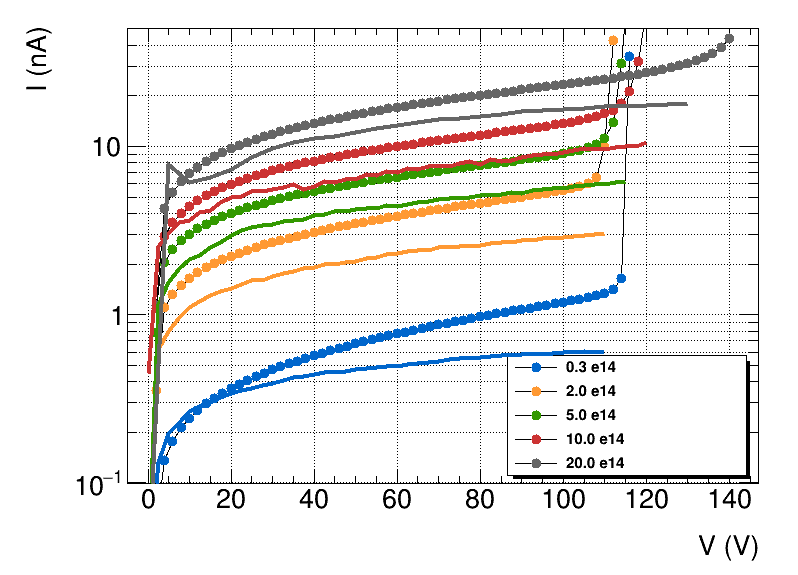} \\ 
   a) & b) \\
 \end{tabular}

\caption{I-V measured after 80 minutes annealing for samples W8 in (a) and W14 in (b). Points show measurements and lines show current values calculated from depletion volume measured with E-TCT and with constant $\alpha$(80 min) = 4$\cdot10^{-17}$ A/cm.}
\label{Current_Measured_Calc}
\end{figure}

Reverse current in the central pixel of the test structure was measured as a function of bias voltage. The bond pads were contacted on in a probe station and the current measured with a Keithley 6517A. The current of the central pixel was measured (see Figure \ref{schemes1}) while the outer pixels were kept at the same potential as the central one forming an effective guard ring. Samples were temperature stabilized at 20$^\circ$C $\pm$ 0.5 $^\circ$C. Figure \ref{Current_Measured_Calc} shows I-V curves for W8 and W14 irradiated to different fluences after annealing for 80 minutes at 60$^\circ$C. 

The standard parametrization of the thermally generated current in the depleted volume in irradiated silicon is  \cite{RD48}: $I = \alpha (t) \cdot V \cdot \Phi_{\mathrm{eq}}$  where $\alpha$(80 min) = 4$\cdot10^{-17}$ A/cm is the value of the constant after annealing for 80 minutes at 60$^\circ$C, $V$ is depleted volume and $\Phi_{\mathrm{eq}}$ equivalent fluence. The depleted volume $V$ can be estimated from E-TCT measurement of the depletion depth $d$ vs. bias voltage shown in Figure \ref{deplvsbias} as $V = d\cdot S$ where $S$ is the pixel surface $60\times 60$ $\mu$m$^2$.  The values of the calculated current can be seen in Figure \ref{Current_Measured_Calc} showing that such simple estimation can describe the measured values to within a factor of 2 over almost two orders of magnitude.

\begin{figure}
\centering
\includegraphics[width=0.6\textwidth]{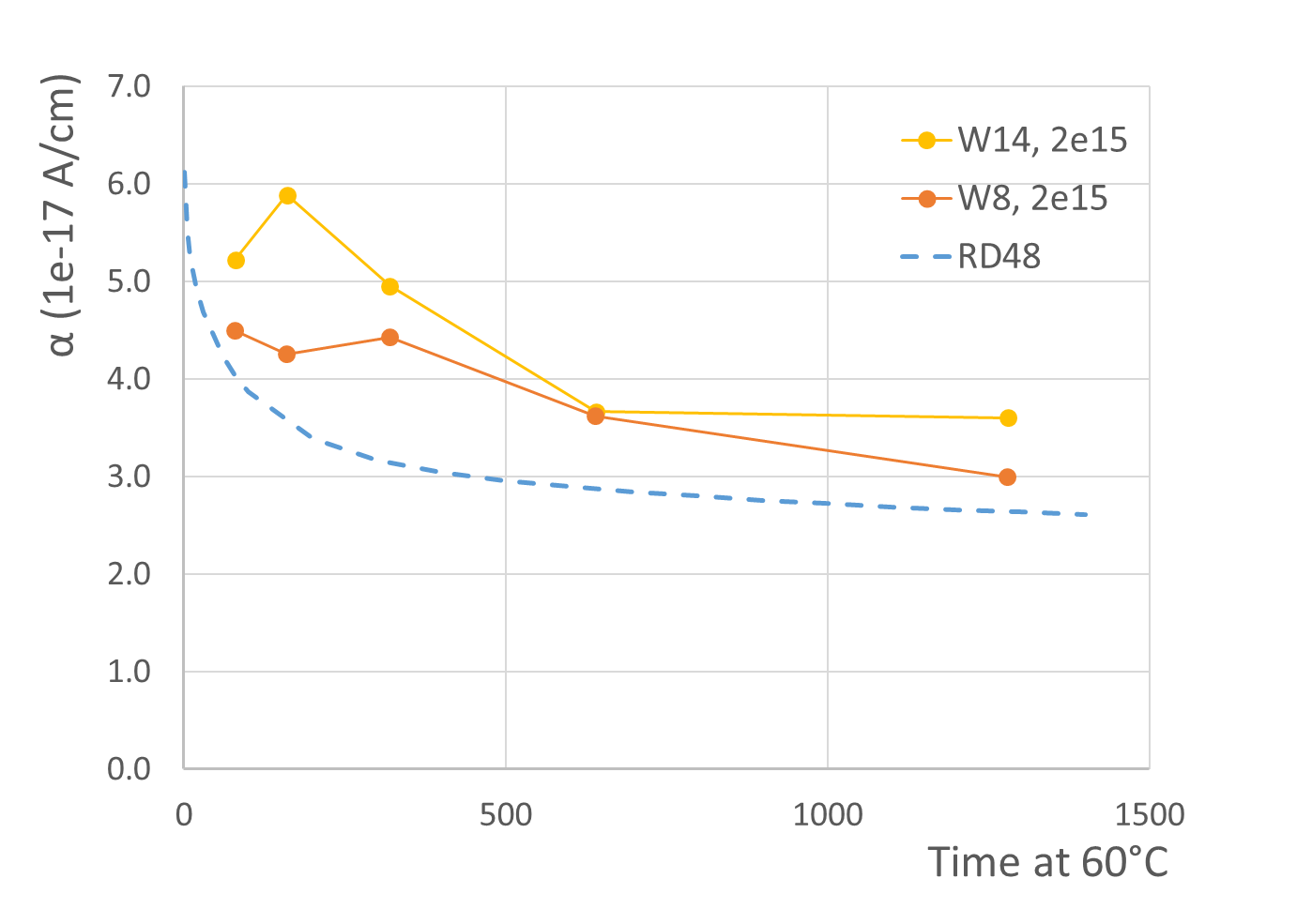} 

\caption{Annealing of current constant $\alpha$ measured with W8 and W14 irradiated to 2$\cdot 10^{15}$n/cm$^2$. The measured value of $\alpha$ is extracted from the curent measured at a bias voltage of 60 V and from the depletion volume measured with E-TCT. Annealing curve of the RD48 model  (\cite{RD48}) is also shown for comparison.}
\label{Current_Measured_Calc_annealing}
\end{figure}

The effect of long term annealing at 60$^\circ$C on the leakage current was studied with samples W8 and W14 irradiated to $\Phi_{\mathrm{eq}} = 2\cdot 10^{15}$ n/cm$^2$. In Figure \ref{Current_Measured_Calc_annealing} the value of parameter $\alpha = I/(V \cdot \Phi_{\mathrm{eq}})$, where $I$ is the current measured at 20$^\circ$C at a bias of 60 V and $V$ is the depleted volume estimated from the E-TCT measurement, is shown as a function of annealing time at 60$^\circ$C. The measured $\alpha$ is compared with the RD48 current annealing model from \cite{RD48}. The decrease of current with increasing annealing time can clearly be seen but the deviations from the model are significant and they originate mainly from a large uncertainty on the estimation of the depleted volume.

\section{Conclusions}

In this paper measurements with irradiated passive pixels on RD50-MPW1 and RD50-MPW2 chips manufactured in LFoundry 150 nm HV-CMOS process are presented. Measurements were made with chips made on wafers with four different resistivities irradiated with reactor neutrons to fluences ranging from 1$\cdot$10$^{13}$ n$_{\mathrm{eq}}$/cm$^2$ to 2$\cdot$10$^{15}$ n$_{\mathrm{eq}}$/cm$^2$. The depletion depth, measured with E-TCT, and leakage current were measured at different bias voltages and after different times of annealing at 60$^\circ$C. 

The effective space charge concentration $N_{\mathrm{eff}}$ was estimated from the dependence of the depletion depth on bias voltage. The observed change of $N_{\mathrm{eff}}$ with neutron fluence was in agreement with previous measurements - an effect of initial acceptor removal was observed on all but the highest initial resistivities. Measurements indicate that the material with higher initial resistivity should be chosen to achieve a larger depletion depth in the whole fluence range.


It was shown that the acceptor removal constant is larger in materials with higher initial resistivities and that stable damage introduction rate is approximately as expected for neutron irradiated silicon. Annealing studies show an expected beneficial effect (i.e. drop of $N_{\mathrm{eff}}$ and consequent increase of depletion depth) of annealing for 80 minutes at 60$^\circ$C. With two samples irradiated to the highest fluence, measurements were made up to a cumulative annealing time of 1280 minutes. No significant long term annealing effect was observed which could be the consequence of the detector material being Cz grown silicon instead of Float Zone.  Significant differences between measured and calculated values of the leakage current as a function of bias voltage, fluence and annealing time were seen but the source of differences are uncertainties of the size of depletion volume. 

With usual model describing irradiated silicon detectors only very rough predictions of the depletion depth and reverse current as a function of fluence, bias voltage and annealing time can be made for investigated devices.

\section{Acknowledgments}

The authors would like to thank the crew at the TRIGA reactor in Ljubljana for help with the irradiation
of the detectors.
Part of this work was performed in the framework of the CERN-RD50 collaboration. The authors acknowledge the financial
support from the Slovenian Research Agency (research core funding No. P1-0135 and project ID PR-06802).

\end{document}